\documentclass[aps, pre, twocolumn]{revtex4-2}

\usepackage[english]{babel}

\usepackage{verbatim}

\usepackage{graphicx}
\usepackage{float}

\usepackage[latin1]{inputenc}

\usepackage{color}

\usepackage[colorlinks,citecolor=blue,linkcolor=black,urlcolor=blue]{hyperref}

\newcommand{\dd}{\mbox{d}}

\newcommand{\eq}[1]{(\ref{#1})}
\newcommand{\bun}{\hat{\mathbf{b}}}

\newcommand{\bv}{\mathbf{v}}
\newcommand{\bB}{\mathbf{B}}
\newcommand{\bJ}{\mathbf{J}}

\newcommand{\dotcross}{ \raise 0.65ex\hbox{${\scriptstyle {{_{\displaystyle \cdot}}\atop\times}}$} }
\newcommand{\crossdot}{ \raise 0.5ex\hbox{${\scriptstyle {{_\times}\atop{\displaystyle \cdot}}}$} }

\newcommand{\cE}{{\cal E}}

\begin{document}

\title{Piecewise omnigenous stellarators with zero bootstrap current}

\date{\today}

\author{Iv\'an Calvo}
\author{Jos\'e Luis Velasco}
\affiliation{Laboratorio Nacional de Fusi\'on, CIEMAT, 28040 Madrid, Spain}
\author{Per Helander}
\affiliation{Max-Planck-Institut f\"ur Plasmaphysik, D-17491 Greifswald, Germany}
\author{F\'elix I. Parra}
\affiliation{Princeton Plasma Physics Laboratory, Princeton, NJ 08540, USA
\vspace{0.5cm}
}

\begin{abstract}
Until now, quasi-isodynamic magnetic fields have been the only known stellarator configurations that, at low  collisionality, give small radial neoclassical transport and zero bootstrap current for arbitrary plasma profiles, the latter facilitating control of the magnetic configuration. The recently introduced notion of piecewise omnigenous fields has enormously broadened the space of stellarator configurations with small radial neoclassical transport. In this Letter, the existence of piecewise omnigenous fields that give zero bootstrap current is proven analytically and confirmed numerically. These results establish piecewise omnigenity as an alternative approach to stellarator reactor design.
\end{abstract}


\maketitle

Stellarators and tokamaks are the leading concepts for magnetic confinement fusion reactors~\cite{Helander2012}. Unlike in tokamaks, where part of the magnetic field is produced by a large electric current induced in the plasma, the confining magnetic field of stellarators is mainly generated by external coils. This enables steady-state operation and prevents current-driven plasma instabilities, two desirable properties in future fusion power plants. These advantages, however, come at the expense of more complicated magnetic configurations. The magnetic field of the tokamak is axisymmetric, but stellarator configurations are intrinsically three-dimensional, something that has important consequences for confinement. Whereas axisymmetry guarantees good confinement of collisionless particle orbits in tokamaks, in stellarators this can only be achieved by means of a careful and sophisticated design of the three-dimensional field, usually referred to as optimization. Without this optimization, stellarator magnetic fields give intolerably large radial neoclassical transport at low collisionality.

Fortunately, the large number of additional degrees of freedom of the stellarator magnetic field make such optimization possible. Typically, stellarator optimization relies on the notion of omnigenity~\cite{Cary1997, Parra2015, Helander2014}. In omnigenous fields, collisionless particles do not move radially on average and, in the course of their motion along magnetic surfaces (also called flux surfaces), they never undergo transitions between different types of orbits. These properties of collisionless orbits are common to omnigenous stellarators and tokamaks, which give similar, very small levels of radial neoclassical transport at the low collisionalities that are characteristic of fusion-grade plasmas. References  \cite{Henneberg_2019, Kinoshita2019, Bader2020, Landreman_2022, Sanchez2023, Goodman2023, Goodman2024, Dudt2024} report on approximately omnigenous fields obtained over the last few years, when stellarator optimization research has witnessed remarkable progress thanks, to a large extent, to the increase of computing power and the development of optimization codes.  Quasi-isodynamic fields, a subset of omnigenous fields that give zero bootstrap current at low collisionality~\cite{Helander2009, Helander2011}, are particularly interesting optimized configurations. The bootstrap current is the net electric current parallel to the magnetic field that emerges in the plasma without current drive. The zero bootstrap current property of quasi-isodynamic fields makes them compatible with an island divertor (one of the most advanced exhaust solutions for stellarators), because even a small bootstrap current can modify the magnetic island structure. Quasi-isodynamicity is the approach Wendelstein 7-X is based on~\cite{Grieger_fst_1992}, and probably the most mature concept for stellarator reactors~\cite{Sanchez2023, Goodman2024, Hegna2025, Lion2025}.

The recent introduction of the notion of piecewise omnigenity~\cite{Velasco2024} represents a major theoretical breakthrough. In piecewise omnigenous fields, the average radial displacement of collisionless particles is zero, but in their motion over magnetic surfaces, particles experience transitions between different types of trapped orbits. The discovery of piecewise omnigenous fields radically expands the space of known configurations that are optimized with respect to radial neoclassical transport. In particular, piecewise omnigenous fields are free from the topological constraints obeyed by magnetic field strength contours in omnigenous configurations, which must close poloidally (this is the case of quasi-isodynamic fields), toroidally or helically, often leading to complex coils. In \cite{Velasco2024}, evidence is provided that piecewise omnigenity may open the door to sufficiently optimized configurations generated by simpler coils.

In this Letter, we show that the potential of piecewise omnigenity goes far beyond the optimization of radial neoclassical transport by proving that there exist piecewise omnigenous fields that give zero bootstrap current at low collisionality for any density and temperature profiles. We will carry out an explicit calculation for a prototypical piecewise omnigenous field and will derive the simple mathematical condition that it has to satisfy to give vanishing bootstrap current. The accuracy of this condition will be confirmed by means of neoclassical simulations. Consequently, two fundamental properties of quasi-isodynamicity (small radial neoclassical transport and bootstrap current) can also be attained by piecewise omnigenity, providing a pathway to new, possibly simpler stellarator reactor designs.

We consider a plasma confined by a time-independent magnetic field $\bB$ tracing out nested toroidal surfaces that we label with a radial coordinate $r$. Ideal magnetohydrodynamic equilibrium determines the component of the electric current $\bJ$ perpendicular to $\bB$. Namely,
\begin{equation}\label{eq:MHD_equilibrium}
\bJ = B^{-1} \bun \times \nabla p + J_{||}\bun,
\end{equation}
where $B = |\bB|$ is the magnetic field strength, $\bun = B^{-1} \bB$ is the unit vector in the direction of the magnetic field and $p(r)$ is the plasma pressure. Imposing $\bJ$ to be divergenceless gives the piece of $J_{||}$ that varies on the flux surface. In the absence of current driven by external sources, the average of $J_{||}$ over the flux surface is the bootstrap current.

The calculation of the bootstrap current at low collisionality is a kinetic problem. We assume that each species $s$ in the plasma is strongly magnetized, $\rho_{s*} := \rho_s / L \ll 1$, so that we can expand the kinetic equation of species $s$ in $\rho_{s*} \ll 1$, average out the fast gyration of particles around magnetic field lines and obtain a dimensionally reduced equation called drift-kinetic equation~\cite{Hazeltine1973b}. Here, $\rho_s = v_{ts}/\Omega_s$ is the Larmor radius of species $s$, $v_{ts}$ is the thermal speed, $\Omega_s = Z_s e B / m_s$ is the Larmor frequency, $Z_s e$ and $m_s$ are, respectively, the electric charge and mass of species $s$, $e$ is the proton charge and $L$ is a characteristic length of variation of $\bB$, on the order of the system size.

As spatial coordinates we will use $\{r, \theta, \zeta\}$, where $\theta$ and $\zeta$ are Boozer angles~\cite{Boozer1981}. In these coordinates, the magnetic field has two particularly useful representations,
$ \bB = \Psi'_t \nabla r \times \left( \nabla\theta - \iota \nabla\zeta \right)$ and
$\bB = I_t \nabla\theta + I_p  \nabla\zeta + \eta \nabla r$, where $2\pi \Psi_t(r)$ is the toroidal flux, $\iota(r)$ is the rotational transform, $I_t(r)$ and $I_p(r)$ are proportional to the toroidal and poloidal currents, respectively, and primes denote derivatives with respect to $r$.

In velocity space, we employ as independent coordinates $\{\cE, \mu, \sigma\}$, where $\cE = v^2/2 + Z_s e \varphi_0/m_s$ is the total energy per mass unit, $\mu = v_\perp^2/2B$ is the magnetic moment and $\sigma = v_{||}/|v_{||}|$ is the sign of the parallel velocity. Here, $v$ is the magnitude of the velocity $\bv$, $v_{||}$ and $v_\perp$ are the components of $\bv$ that are, respectively, parallel and perpendicular to $\bB$, and $\varphi \simeq \varphi_0(r)$ is the lowest-order electrostatic potential in an expansion in $\rho_{s*} \ll 1$.

The drift-kinetic equation for species $s$ reads
\begin{equation}
v_{||}\bun\cdot\nabla h_s + \bv_{Ms}\cdot\nabla r \, \partial_r f_{Ms} = \sum_{s'} C_{ss'}^{(l)}[h_s;h_{s'}], 
\end{equation}
where $h_s(r, \theta,\zeta,\cE,\mu,\sigma)$ is the non-adiabatic component of the deviation of the distribution function from a Maxwellian
\begin{equation}
f_{Ms} (r,\cE) = n_s\left(\frac{m_s}{2\pi T}\right)^{3/2}\exp\left(- \frac{m_s\cE - Z_s e \varphi_0}{T}\right),
\end{equation}
the density $n_s(r)$ and temperature $T(r)$ are flux functions,
\begin{equation}
\bv_{Ms}\cdot\nabla r = \frac{v_{||}}{B} (\bB\times\nabla r)\cdot \nabla\left(\frac{v_{||}}{\Omega_s}\right)
\end{equation}
is the radial drift due to the spatial variation of the magnetic field and $C_{ss'}^{(l)}$ is the linearized Landau collision operator~\cite{Landau1936}, whose explicit form is not needed here. The expression of the parallel velocity in terms of the independent coordinates is $v_{||} = \sigma \sqrt{2(\cE - Z_s e \varphi_0/m_s - \mu B)}$, which depends on $\theta$ and $\zeta$ through $B(r,\theta,\zeta)$.

In a piecewise omnigenous field, all particles have vanishing average radial magnetic drift. Hence, there exist functions $F_s(r, \theta,\zeta,\cE,\mu,\sigma)$ such that
\begin{equation}\label{eq:def_Fs}
v_{||}\bun\cdot \nabla F_s
=
-
\bv_{Ms}\cdot\nabla r \, \partial_r f_{Ms}.
\end{equation}
Defining $g_s : = h_s - F_s$, the drift-kinetic equation can be written as
\begin{equation}
v_{||}\bun\cdot\nabla g_s = \sum_{s'} C_{ss'}^{(l)}[g_s + F_s; g_{s'} + F_{s'}].
\end{equation}
We assume that all species are in the banana regime and expand in small collisionality $\nu_{s*} \ll 1$ (see e.g.~\cite{Helander2009}). Writing $g_s = g_{s0} + g_{s1} + \dots$, where $g_{sk} = O(\nu_{s*}^k \rho_{s*} f_{Ms})$, the lowest-order drift-kinetic equation gives
\begin{equation}\label{eq:lowest_order_DKE}
v_{||}\bun\cdot\nabla g_{s0} = 0.
\end{equation}
We need to distinguish between passing trajectories, where $v_{||}$ never vanishes, and trapped trajectories, where $v_{||}$ vanishes at two points called bounce points. Passing particles are characterized by $ 0 \le \lambda: = \mu/ (\cE - Z_s e \varphi_0) < B_M^{-1}$ and trapped particles by $B_M^{-1} \le \lambda \le B_m^{-1}$, where $B_M(r)$ and $B_m(r)$ are, respectively, the maximum and minimum values of $B$ on the flux surface $r$. Assuming that the surface is ergodic, equation \eq{eq:lowest_order_DKE} implies $g_{s0} \equiv g_{s0}(r,\cE,\mu,\sigma)$ for passing particles. For trapped particles, it implies that $g_{s0}$ is constant along field lines and even in $\sigma$.

The function $g_{s0}$ is found from the orbit average of the next-to-lowest order drift-kinetic equation. Denote by $\overline{(\cdot)}$ the orbit average operation, defined as the annihilator of $v_{||}\bun\cdot\nabla$. Then, the equation that determines $g_{s0}$ is
\begin{equation}\label{eq:eq_for_g0_general}
\overline{\sum_{s'} C_{ss'}^{(l)}[g_{s0} + F_s; g_{s'0} + F_{s'}]} = 0.
\end{equation}
Since $F_s$ is odd in $\sigma$ and $C_{ss'}^{(l)}$ preserves parity, $g_{s0}$ must be odd in $\sigma$. Above we have argued that, for trapped particles, $g_{s0}$ is even in $\sigma$; then, $g_{s0} \equiv 0$ for trapped particles~\footnote{Along the Letter, when solving equations like \eq{eq:eq_for_g0_general}, we ignore functions that are in the kernel of $C_{ss'}^{(l)}$. These functions can be set to zero by fixing the values of the density, parallel flow and temperature of the Maxwellian distributions, $f_{Ms}$.}. For passing trajectories, the orbit average becomes $\overline{(\cdot)} = \langle B |v_{||}|^{-1} (\cdot)\rangle$, and $g_{s0}$ for passing particles is found from
\begin{eqnarray}\label{eq:eq_for_g0_passing}
&&\left\langle
\frac{B}{|v_{||}|} \sum_{s'} C_{ss'}^{(l)}[g_{s0}; g_{s'0}] 
\right\rangle =
\nonumber\\[5pt]
&&\hspace{1cm}
-
\left\langle
\frac{B}{|v_{||}|} \sum_{s'} C_{ss'}^{(l)}[F_s; F_{s'}] 
\right\rangle,
\end{eqnarray}
where $\langle (\cdot) \rangle$ is the flux surface average~\cite{Helander2014}. Note that both the passing- and trapped-particle components of $F_s$ enter this equation because the linearized Landau collision operator is a differential operator in its first argument and an integral operator in its second argument. The manipulations leading to \eq{eq:eq_for_g0_passing} are valid for a tokamak, an omnigenous stellarator and a piecewise omnigenous stellarator. However, the specific form of $F_s$ in piecewise omnigenous fields substantially differs from the other two cases.

\begin{figure}[h!]
\centering
\includegraphics[width=0.9\columnwidth]{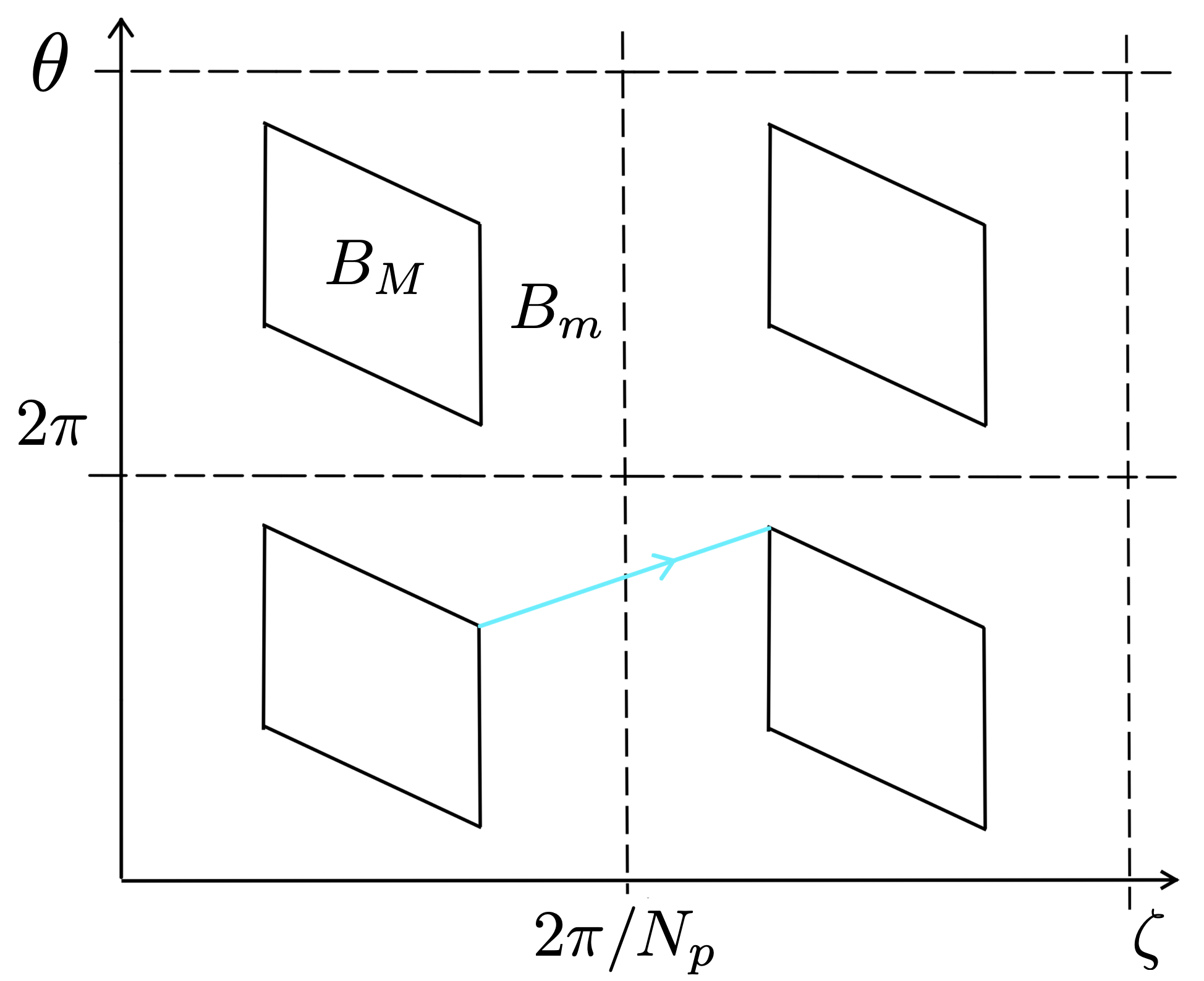}
\caption{$B$ on a flux surface in an example of a prototypical piecewise omnigenous field. A segment of a field line is drawn in blue. $N_p$ is the number of field periods.}
\label{fig:prototypical_pwO_field}
\end{figure}

In figure \ref{fig:prototypical_pwO_field}, we present an example of a prototypical piecewise omnigenous field. The magnetic field strength $B$ takes only two values on each flux surface, $B_M$ inside the parallelograms and $B_m$ outside. The value of the rotational transform must ensure that field lines in Boozer coordinates join corners of the parallelograms as shown in figure \ref{fig:prototypical_pwO_field}. The parameterization of these prototypical piecewise omnigenous fields (i.e. the definition of parameters that control the slopes of the sides of the parallelograms, their areas, etc.) is given in \cite{Velasco2024}. If $B$ is one of these prototypical piecewise omnigenous fields, we will be able to provide an explicit expression for $F_s$ and, finally, for the condition for zero bootstrap current. This does not mean that all piecewise omnigenous fields are like the field of figure \ref{fig:prototypical_pwO_field}, which should be viewed as a useful, simple and limit case. In \cite{Velasco2025}, much more general constructions are given. Examples of recently obtained approximately piecewise omnigenous fields that satisfy the ideal magnetohydrodynamic equilibrium equation can be found in \cite{Bindel2023, Liu_arXiv_2025}.

\begin{figure}
\centering
\includegraphics[width=0.9\columnwidth]{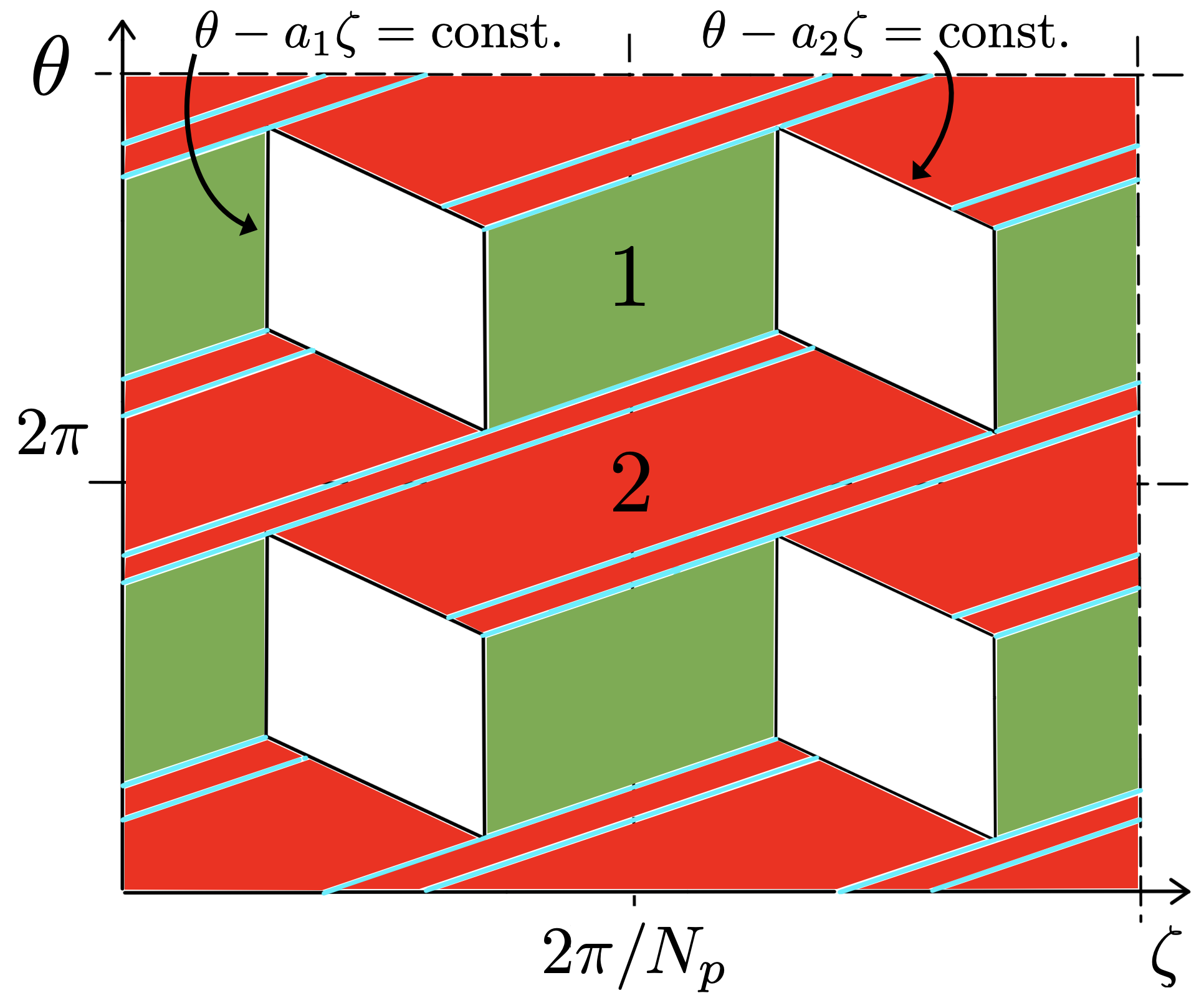}
\caption{Definition of regions 1 (green) and 2 (red), where $F_s$ takes its two different non-zero values. Inside the parallelograms (white), $F_s \equiv 0$.}
\label{fig:regions_1_and_2}
\end{figure}

The calculation of $F_s$ in a prototypical piecewise omnigenous field is the most technical part of the derivation of the result of this Letter and is carried out in detail in Appendix \ref{sec:appendix}. Next, we give the expression for $F_s$. On each flux surface and for fixed values of the velocity coordinates, $F_s$ is piecewise constant. Inside the parallelograms, $F_s \equiv 0$. Outside the parallelograms, we define regions 1 and 2 as shown in figure \ref{fig:regions_1_and_2}. In each of these two regions, $F_s$ is constant  and, in general, non-zero. We say that a side of a parallelogram is of type $i$ if it is part of the boundary of region $i$, and parameterize the sides of the parallelograms as $\theta - a_i \zeta = \mbox{const.}$, where $a_i$ is the slope of sides of type $i$. We can give a compact expression for $F_s$. For passing particles, $\lambda < B_M^{-1}$,
\begin{equation}\label{eq:Fs_passing}
F_s = \sum_i
 \left(
 \frac{v_{||}}{\Omega_s}\Big\vert_M - \frac{v_{||}}{\Omega_s}\Big\vert_m
 \right)
 \frac{1}{\Psi'_t} \, \frac{a_i I_t + I_p}{\iota -a_i} \, \partial_r f_{Ms} \, \chi_i.
\end{equation}
For trapped particles, $\lambda \ge B_M^{-1}$,
\begin{eqnarray}\label{eq:Fs_trapped}
F_s &=&
\left(
\frac{v_{||}}{\Omega_s}\Big\vert_m - \frac{v_{||}}{\Omega_s}\Big\vert^{\lambda = B_M^{-1}}_m \right)
\frac{I_t}{\Psi'_t} \,
\partial_r f_{Ms}
\nonumber\\[5pt]
&-&
 \sum_i
  \frac{v_{||}}{\Omega_s}\Big\vert^{\lambda = B_M^{-1}}_m
  \frac{1}{\Psi'_t} \, \frac{a_i I_t + I_p}{\iota -a_i} \, \partial_r f_{Ms} \, \chi_i.
\end{eqnarray}
Here, $\chi_i(r,\theta,\zeta)$ is the characteristic function of region $i$, so that $\chi_i = 1$ in region $i$ and $0$ otherwise, $(\cdot)\vert_M$ and $(\cdot)\vert_m$ stand for the evaluation of $(\cdot)$ at $B = B_M$ and $B = B_m$, respectively, and $(\cdot)\vert^{\lambda = B_M^{-1}}$ denotes evaluation at the value of $\lambda$ corresponding to the interface between passing and trapped particles. Note that \eq{eq:Fs_passing} and \eq{eq:Fs_trapped} imply that $F_s$ is continuous at $\lambda = B_M^{-1}$.

Although we will rigorously show it below, we can intuitively understand how a piecewise omnigenous field can give zero bootstrap current by inspection of the expression for $F_s$. If $\iota - a_1$ and $\iota - a_2$ have different signs, then $F_s$ has opposite signs in regions 1 and 2, and the contributions from these two regions can, in principle, cancel out. We proceed to give the mathematical proof and derive the precise condition that $B$ has to satisfy.

Recall equation \eq{eq:eq_for_g0_passing}, which determines $g_{s0}$ for passing particles, and focus on its right-hand side. $C_{ss'}^{(l)}$ is an integro-differential operator in coordinates $\cE$ and $\mu$, with coefficients that depend on space through $B$. These coefficients are constant outside the parallelograms and $F_s \equiv 0$ inside. Using this and expressions \eq{eq:Fs_passing} and \eq{eq:Fs_trapped}, we infer that \eq{eq:eq_for_g0_passing} can be written as
\begin{equation}
\left\langle
\frac{B}{|v_{||}|} \sum_{s'} C_{ss'}^{(l)}[g_{s0}; g_{s'0}] 
\right\rangle
= 
I_t \Lambda_{t, s} + I_p \Delta \Lambda_{p,s},
\end{equation}
where $\Lambda_{t,s} (r,\cE,\mu,\sigma)$ and $\Lambda_{p,s} (r,\cE,\mu,\sigma)$ are phase-space functions whose explicit form we do not need and
\begin{equation}\label{eq:def_Delta}
\Delta := \frac{\left\langle \chi_1 \right\rangle}{\iota - a_1} + \frac{\left\langle \chi_2 \right\rangle}{\iota - a_2} \, 
\end{equation}
is a geometric quantity that will be key in what follows. Analogously, for some phase-space functions $\Lambda^g_{t,s}$ and $\Lambda^g_{p,s}$, and some flux functions $k^V_{t,s}$, $k^V_{p,s}$, $k^J_t$ and $k^J_p$, we can express $g_{0s}$, the net parallel flow of each species $\langle V_{||,s} B \rangle := \left\langle B \int v_{||} \left( g_{s0} + F_s \right) \dd^3 v \right\rangle$ and the bootstrap current $\langle J_{||} B \rangle = \sum_s Z_s e \langle V_{||,s} B \rangle $ as
\begin{equation}
g_{0s} = I_t \Lambda^g_{t, s} + I_p \Delta \Lambda^g_{p,s},
\end{equation}
\begin{equation}\label{eq:expression_Vpar}
\langle
V_{||,s} B
\rangle
=
I_t k^V_{t, s} + I_p \Delta k^V_{p,s}
\end{equation}
and
\begin{equation}\label{eq:expression_Jpar}
\langle
J_{||} B
\rangle
= I_t k^J_t + I_p \Delta k^J_p.
\end{equation}

We are ready to ask the question that is the subject of this Letter: \emph{Is there a subset of piecewise omnigenous fields that give $\left\langle J_{||} B \right\rangle \equiv 0$ for arbitrary plasma profile gradients?} In order to answer the question, we invoke a consistency condition between $I_t$ and $\langle J_{||} B \rangle$ imposed by ideal magnetohydrodynamic equilibrium. Taking the toroidal component of \eq{eq:MHD_equilibrium} and after easy manipulations (see e.g. \cite{Landreman2012}), one obtains the differential equation
\begin{equation}\label{eq:consistency_condition_from_MHD}
I_t' + \frac{\mu_0 p'}{\langle B^2 \rangle} I_t = \frac{\mu_0}{\langle B^2 \rangle} \langle J_{||} B \rangle,
\end{equation}
where $\mu_0$ is the vacuum permeability. This equation is general. If the expression of the bootstrap current turns out to be such that $\langle J_{||} B \rangle \propto I_t$, then equation \eq{eq:consistency_condition_from_MHD} and the boundary condition $I_t (r = 0) = 0$ imply $I_t \equiv 0$ in the whole plasma volume and, therefore, $\left\langle J_{||} B \right\rangle \equiv 0$. For example, the subset of omnigenous fields that have $\langle J_{||} B \rangle \propto I_t$ (hence, vanishing bootstrap current) are fields with poloidally closed $B$ contours, i.e. quasi-isodynamic fields~\cite{Helander2009}. In the case of the piecewise omnigenous field considered in this Letter, the condition for zero bootstrap current is quite different. From equation \eq{eq:expression_Jpar}, we deduce that the condition is
\begin{equation}\label{eq:Delta_equals_zero}
\Delta = 0.
\end{equation}
We point out that condition \eq{eq:Delta_equals_zero} actually implies that $\langle V_{||,s} B \rangle$ vanishes for each species. This is analogous to the case of omnigenous fields, for which poloidally closed $B$ contours imply zero net parallel flow for all species, not just zero bootstrap current~\cite{Helander2009}.

\begin{figure}[h!]
\centering
\includegraphics[width=\columnwidth]{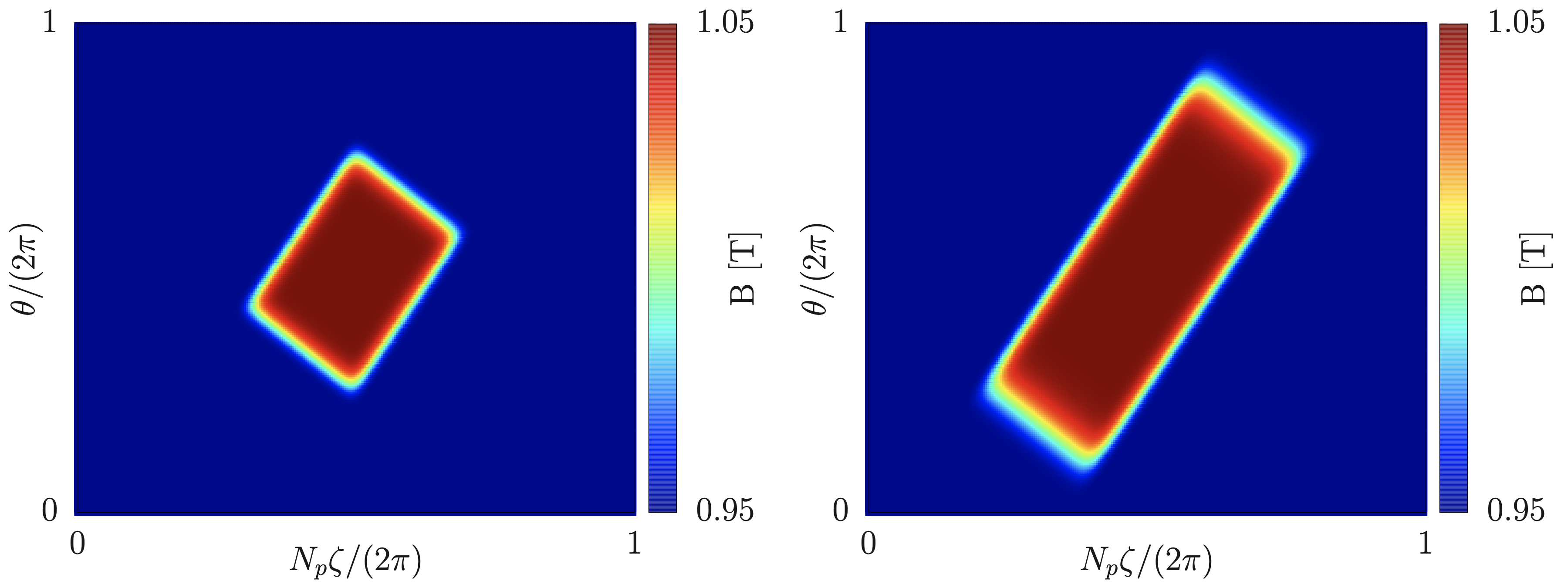}
\caption{$B$ obtained by using expression \eq{eq:B_pwO_parameterization} for $w_2 = 0.5\pi$ (left) and $w_2 = \pi$ (right). The remaining values of the parameters are $q = 6$, $N_p = 2$, $w_1=0.3\pi/N_p$, $\theta_c=\pi$, $\zeta_c=\pi/N_p$, $t_1= - 0.3$, $t_2 = 1.8$, $\iota = 0.435$, $B_m = 0.95$ and $B_M = 1.05$. The values of $\Delta$ are $\Delta = 0.093$ (left) and $\Delta = 0$ (right).}
\label{fig:B_for_two_w2_values}
\end{figure}

Finally, we confirm the analytical result by means of neoclassical simulations with the code \texttt{MONKES}~\cite{Escoto2024}. In order to perform scans in $\Delta$ within the space of piecewise omnigenous fields, we use the parameterization of prototypical piecewise omnigenous fields introduced in \cite{Velasco2024} and given by
\begin{eqnarray}\label{eq:B_pwO_parameterization}
&&\hspace{-0.7cm}\frac{B - B_m}{B_M - B_m} = 
e^{
-\left(\frac{\zeta - \zeta_c + t_1(\theta - \theta_c)}{w_1}\right)^{2q}
-\left(\frac{\theta - \theta_c + t_2(\zeta - \zeta_c)}{w_2}\right)^{2q}
}
\end{eqnarray}
and
\begin{equation}
\iota =
\left( \frac{\pi (1 - t_1 t_2)}{N_p w_1} -1 \right)^{-1} t_2,
\end{equation}
where $q$, $\zeta_c$, $\theta_c$, $t_1$, $t_2$, $w_1$ and $w_2$ are free parameters. These magnetic fields are exactly piecewise omnigenous only if $q\to \infty$. In this limit, $\Delta \propto (\pi - w_2)$, whereas $\iota$ is independent of $w_2$ for every $q$. Hence, varying $w_2$ while keeping the values of the remaining parameters constant is a convenient way of changing $\Delta$.  In figure \ref{fig:B_for_two_w2_values}, we plot the $B$ given by \eq{eq:B_pwO_parameterization} for two sets of values of the parameters, that will also be employed in the scans below.
  
\begin{figure}[h!]
\centering
\includegraphics[width=\columnwidth]{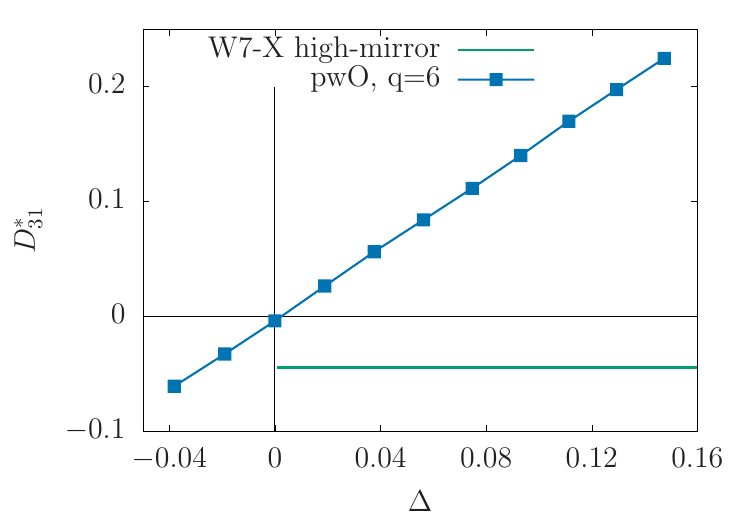}
\caption{$D_{31}^*$ as a function of $\Delta$ for $I_t = 0$, $I_p \neq 0$ and $\nu^* = 7\cdot10^{-4}$. The normalizations employed in $D_{31}^*$ and the collisionality $\nu^*$ are defined in \cite{Beidler2011}. The values of $q$, $N_p$, $w_1$, $\theta_c$, $\zeta_c$, $t_1$, $t_2$, $\iota$, $B_m$ and $B_M$, appearing in expression \eq{eq:B_pwO_parameterization} for approximately piecewise omnigenous (pwO) fields, are the same as in figure \ref{fig:B_for_two_w2_values}. For reference, the value of $D_{31}^*$ in the high-mirror configuration of Wendelstein 7-X is shown.}
\label{fig:d31_vs_delta}
\end{figure}

\begin{figure}[h!]
\centering
\includegraphics[width=\columnwidth]{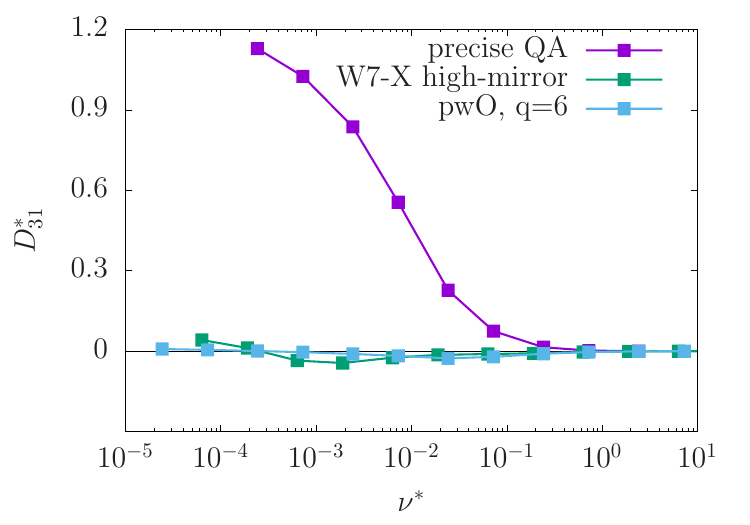}
\caption{$D_{31}^*$ as a function of $\nu^*$ for $\Delta = 0$. The values of the remaining parameters are the same as in figure \ref{fig:d31_vs_delta}. For reference, the values of $D_{31}^*$ in the precise QA configuration of \cite{Landreman_2022} and the high-mirror configuration of Wendelstein 7-X are shown.}
\label{fig:d31_vs_nu}
\end{figure}

\begin{figure}[h!]
\centering
\includegraphics[width=\columnwidth]{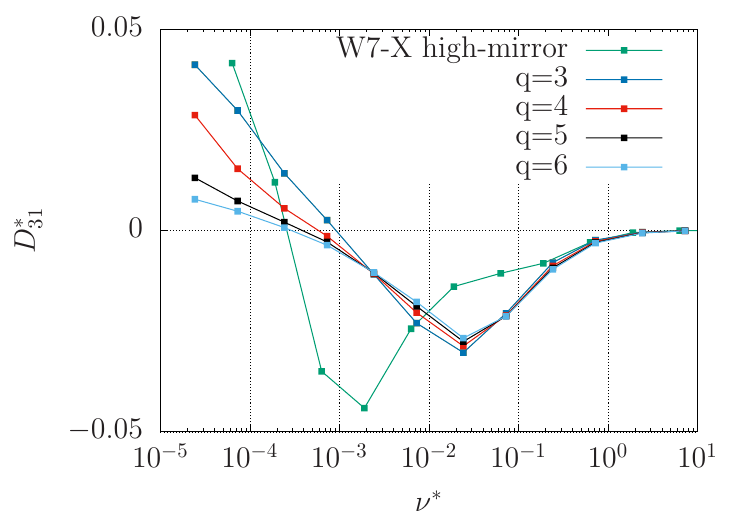}
\caption{$D_{31}^*$ as a function of $\nu^*$ for $\Delta = 0$ and for several values of $q$. The values of the remaining parameters are the same as in figure \ref{fig:d31_vs_nu}. For reference, the values of $D_{31}^*$ in the high-mirror configuration of Wendelstein 7-X are shown.}
\label{fig:d31_vs_q}
\end{figure}

The main numerical result of the Letter is contained in figure \ref{fig:d31_vs_delta}, where the monoenergetic transport coefficient $D_{31}$ (which gives the boostrap current, see e.g. \cite{Beidler2011}) is calculated as a function of $\Delta$. The analytical result is confirmed, with $D_{31}$ vanishing at $\Delta = 0$. In figure \ref{fig:d31_vs_nu}, we plot $D_{31}$ for $\Delta = 0$ as a function of the collisionality, showing that the result holds in a broad range of low collisionalities. In figure \ref{fig:d31_vs_q}, zoomed-in views of the $q=6$ and W7-X curves are given, together with results for smaller values of $q$. Even for moderately small values of $q$, $D_{31}^*$ at low collisionality is smaller than in W7-X.
 
In summary, we have proven that piecewise omnigenous fields, recently introduced as a new class of magnetic fields optimized for radial neoclassical transport, can also be optimized to give zero bootstrap current for any density and temperature profiles. Therefore, the two main physical properties of quasi-isodynamic fields can also be achieved with piecewise omnigenous fields. This consolidates piecewise omnigenity as a novel approach to the design of stellarator reactors.

\begin{acknowledgments}
This work has been carried out within the framework of the EUROfusion Consortium, funded by the European Union via the Euratom Research and Training Programme (Grant Agreement No 101052200 -- EUROfusion). Views and opinions expressed are however those of the author(s) only and do not necessarily reflect those of the European Union or the European Commission. Neither the European Union nor the European Commission can be held responsible for them. This work was supported by the U.S. Department of Energy under contract number DE-AC02-09CH11466. The United States Government retains a non-exclusive, paid-up, irrevocable, world-wide license to publish or reproduce the published form of this manuscript, or allow others to do so, for United States Government purposes. This research was supported in part by Grant No. PID2021-123175NB-I00, funded by Ministerio de Ciencia, Innovaci\'on y Universidades / Agencia Estatal de Investigaci\'on / 10.13039/501100011033 and by ERDF/EU. The authors would like to thank F.~J. Escoto for his advice on the use of \texttt{MONKES}.
\end{acknowledgments}

\vspace{0.5cm}

The data that support the findings of this article are openly available \cite{data}.

\appendix

\section{Calculation of $F_s$}
\label{sec:appendix}

\begin{figure}[h!]
\centering
\includegraphics[width=0.7\columnwidth]{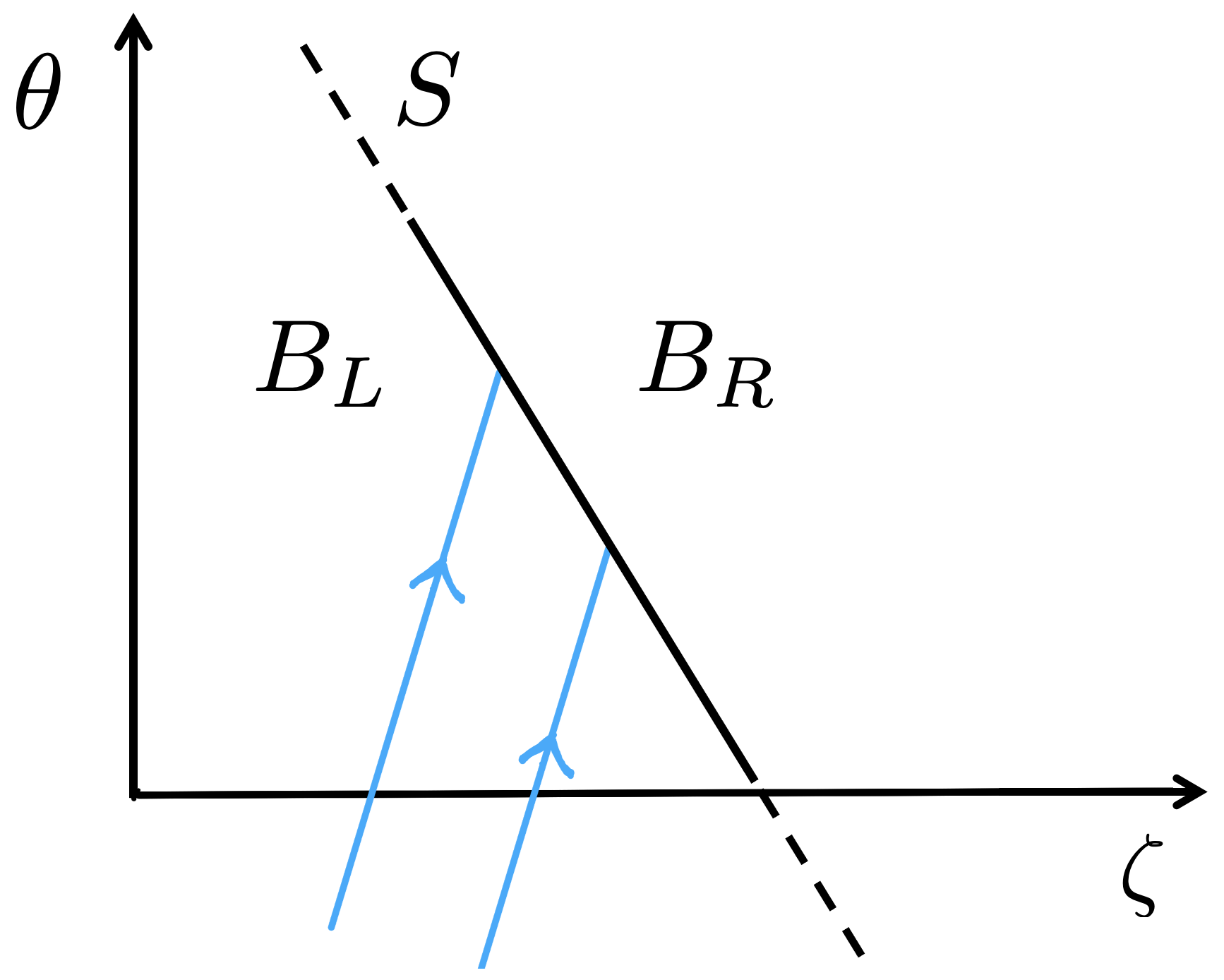}
\caption{The basic result in order to find $F_s$ for passing particles consists of the calculation of the jump of $F_s$ along a straight line separating two regions where $B$ takes two different values.}
\label{fig:building_block}
\end{figure}

Here, we explain how to obtain \eq{eq:Fs_passing} and \eq{eq:Fs_trapped} from \eq{eq:def_Fs} when $B$ is a prototypical piecewise omnigenous field. We start with passing particles. The source term of the equation for $F_s$ vanishes in regions of constant $B$. Hence, $F_s$ is constant except at points where $B$ jumps; i.e. except at the boundary of the parallelograms. We will work out the jump of $F_s$ at points on a straight line $S$ separating a region with $B = B_L$ from a region with $B = B_R$ (see figure \ref{fig:building_block} for a sketch). Then, we will apply the result to the case where $S$ is (locally) one of the sides of the parallelograms in our piecewise omnigenous field, and $B_L$ and $B_R$ stand for either $B_M$ or $B_m$. In figure \ref{fig:building_block} and in the calculation below, we assume  
$\iota - a > 0$, but the final result is valid also for $\iota -a < 0$. 

Let $S$ be parameterized by $\theta = a \zeta + c$. In Boozer coordinates, the differential equation for $F_s$ reads
\begin{eqnarray}\label{eq:eq_Fs_Boozer_coor}
&&\partial_\zeta F_s + \iota \partial_\theta F_s =
\\[5pt]
&&\hspace{0.3cm}
\left(\frac{v_{||}}{\Omega_s}\Big\vert_L - \frac{v_{||}}{\Omega_s}\Big\vert_R\right)
\frac{1}{\Psi'_t} ( a I_t + I_p) \, \partial_r f_{Ms} \delta\left( \theta - a\zeta - c \right),
\nonumber
\end{eqnarray}
where $(\cdot)\vert_L$ and $(\cdot)\vert_R$ stand for the evaluation of $(\cdot)$ at $B = B_L$ and $B = B_R$, respectively, and $\delta$ is the Dirac delta function. The function $F_s$ is constant on each side of $S$, with values $F_{s, L}$ and $F_{s, R}$. In order to calculate the jump $F_{s, R} - F_{s, L}$, it is useful to write \eq{eq:eq_Fs_Boozer_coor} in coordinates $\alpha := \theta - \iota \zeta$ and $\zeta$. Namely,
\begin{eqnarray}\label{eq:eq_Fs_alpha_zeta_coor}
&&\partial_\zeta F_s =
\\[5pt]
&&\hspace{0.3cm}
\left(\frac{v_{||}}{\Omega_s}\Big\vert_L - \frac{v_{||}}{\Omega_s}\Big\vert_R\right)
\frac{1}{\Psi'_t} \frac{a I_t + I_p}{\iota - a} \, \partial_r f_{Ms} \delta\left( \zeta + \frac{\alpha - c}{\iota - a} \right),
\nonumber
\end{eqnarray}
where the derivative with respect to $\zeta$ is taken holding $\alpha$ constant. It is straightforward to integrate  \eq{eq:eq_Fs_alpha_zeta_coor} across $S$ and find that
\begin{equation}
F_{s, R} - F_{s, L} =
\left(\frac{v_{||}}{\Omega_s}\Big\vert_L - \frac{v_{||}}{\Omega_s}\Big\vert_R\right)
\frac{1}{\Psi'_t}\frac{a I_t + I_p}{\iota - a} \, \partial_r f_{Ms}.
\end{equation}

\begin{figure}
\centering
\includegraphics[width=0.9\columnwidth]{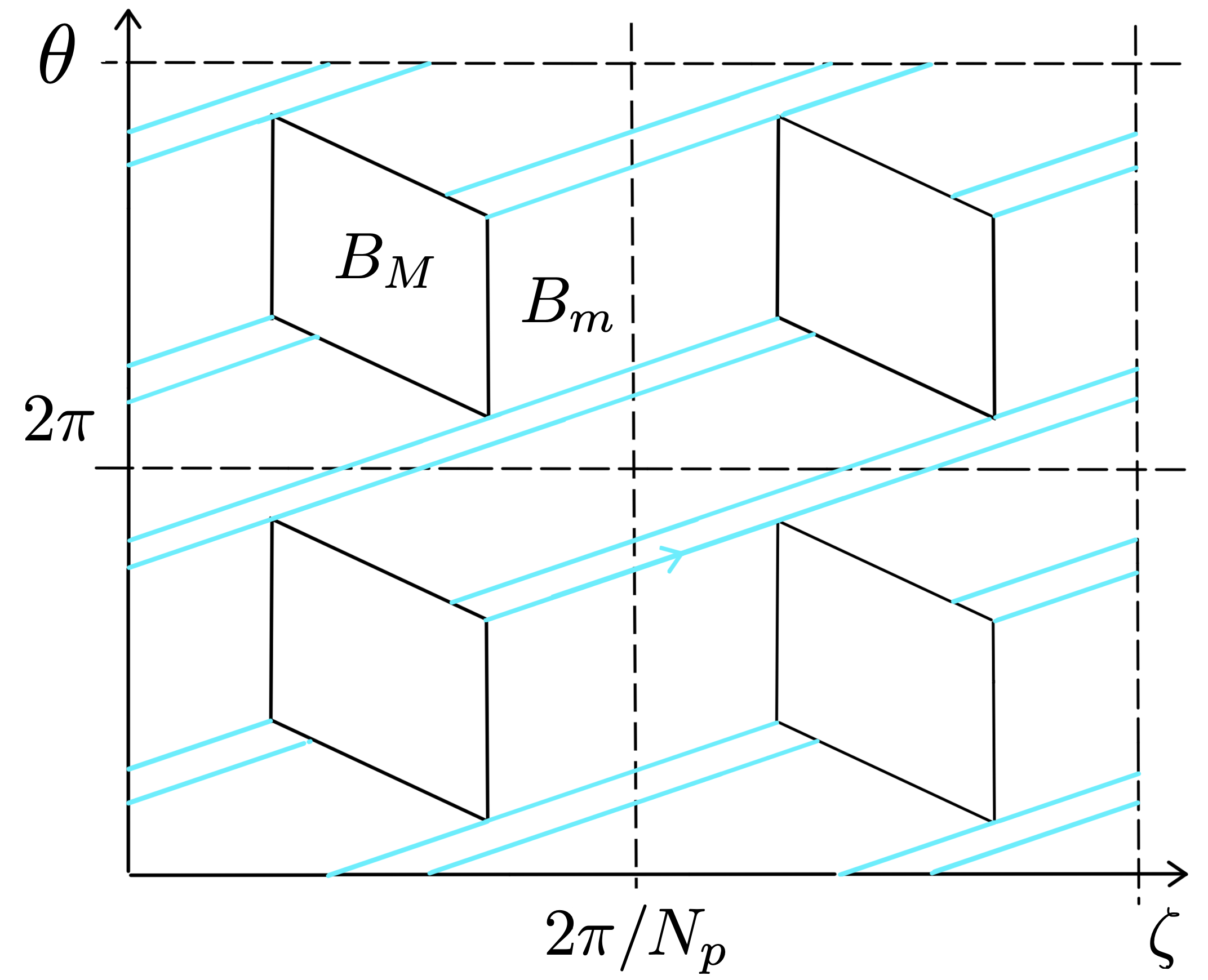}
\caption{This is the result of adding to figure \ref{fig:prototypical_pwO_field} segments of all field lines that hit the corners of the parallelograms.}
\label{fig:B_pwO_model_many_field_lines}
\end{figure}

Next, we apply the above result to the case where $S$ is locally a side of a parallelogram of our piecewise omnigenous field. The parallelograms have two types of sides, depending on their slope. By inspection of figure \ref{fig:B_pwO_model_many_field_lines}, we infer that if a field line leaves a parallelogram through a side of a certain type, then the next time it enters a parallelogram it will be through a side of the same type. Hence, when exiting and then entering a parallelogram, $F_s$ undergoes a jump of the same size and opposite sign. This implies that inside the parallelograms we can choose $F_s \equiv 0$. Outside the parallelograms, we can define two subregions, labeled by $1$ and $2$, where $F_s$ takes the values
\begin{equation}
F_{s, i} =
\left(\frac{v_{||}}{\Omega_s}\Big\vert_M - \frac{v_{||}}{\Omega_s}\Big\vert_m\right)
 \frac{1}{\Psi'_t} \, \frac{a_i I_t + I_p}{\iota -a_i} \, \partial_r f_{Ms}.
\nonumber
\end{equation}
Hence, we have derived expression \eq{eq:Fs_passing}.

We turn to trapped particles. Equation \eq{eq:def_Fs} can be solved as long as all trapped trajectories have zero orbit-averaged radial magnetic drift. In coordinates $\alpha$ and $\zeta$, equation \eq{eq:def_Fs} becomes
\begin{equation}
\partial_\zeta F_s = \frac{1}{\Psi'_t}\left[I_t \partial_\zeta - (I_p + \iota I_t) \partial_\alpha \right]\frac{v_{||}}{\Omega_s} \partial_r f_{Ms}.
\end{equation}
We can integrate this equation to obtain
\begin{eqnarray}\label{eq:Fs_trapped_general}
&& F_s = \frac{I_t}{\Psi'_t} \frac{v_{||}}{\Omega_s} \partial_r f_{Ms}
- \frac{I_p + \iota I_t}{\Psi'_t} \partial_r f_{Ms} \partial_\alpha \int^\zeta \frac{v_{||}}{\Omega_s} \dd\tilde\zeta
\nonumber\\[5pt]
&&\hspace{1cm}
+ K(r,\alpha, \cE,\mu,\sigma),
\end{eqnarray}
where $K$ is an integration constant and the function under the integral depends on $\alpha$ and $\zeta$. In prototypical piecewise omnigenous fields, the second term on the right-hand side of \eq{eq:Fs_trapped_general} is zero because nothing depends on $\alpha$. Then,
\begin{equation}\label{eq:Fs_trapped_pwO}
F_s = \frac{I_t}{\Psi'_t} \frac{v_{||}}{\Omega_s} \partial_r f_{Ms}
+ K.
\end{equation}
In general, we cannot replace $v_{||}/\Omega_s$ by $v_{||}/\Omega_s \vert_m$ in \eq{eq:Fs_trapped_pwO}. For example, in order to understand $F_s$ as the solution to the differential equation \eq{eq:def_Fs}, one should view $B$ as a smooth function on the flux surface that tends to the magnetic field of interest as a limit case; in particular, to a field with square wells. One would then take the limit only at the end of the manipulations. However, for the arguments of this Letter, we can safely take this limit in \eq{eq:Fs_trapped_pwO}. Finally, choosing $K$ to make $F_s$ continuous at $\lambda = B_M^{-1}$, we obtain expression \eq{eq:Fs_trapped}.

\providecommand{\noopsort}[1]{}\providecommand{\singleletter}[1]{#1}%

\end{document}